\documentstyle[psfig]{elsart}

\begin{document}
\begin{frontmatter}
\title{Squeezed condensate of gluons and the mass of the $\eta'$}

\author[rostock]{H.-P. Pavel\thanksref{DFG},}
\author[rostock]{D. Blaschke,}
\author[dubna]{V.N. Pervushin,}
\author[rostock,dubna]{ G. R\"opke} and 
\author[dubna]{M.K. Volkov\thanksref{RFFI}}
\address[rostock]{Fachbereich Physik, Universit\"at Rostock, 
  D-18051 Rostock, Germany}
\address[dubna]{Bogoliubov  Laboratory of Theoretical Physics,\\
       Joint Institute for Nuclear Research, 141980, Dubna, Russia}
\thanks[DFG]{Supported by DFG grant No RO 905/11-1}
\thanks[RFFI]{Supported by INTAS grant No. W 94-2915}

\begin{abstract}
The relation between the large mass of the $\eta'$ and the structure of 
the gluon vacuum via the $U_A(1)$ anomaly is discussed.
A squeezed gluon vacuum is considered
as an alternative to existing models.
Considering Witten's formula for the $\eta_0$ mass
we show that the contact term can give a sizable contribution and relate it
to the physical gluon condensate.
The values of the gluon condensate obtained through this relation are
compared with the value by Shifman, Vainshtein and Zakharov
and the recent update values by Narison.

\vspace{5mm}
\noindent
PACS number(s):  12.38.Aw, 12.40.Yx, 14.40.Aq, 14.70.Dj
\end{abstract}
\end{frontmatter}

\section{Introduction}

In the nonet of pseudoscalar mesons the most interesting one 
for the investigation of the gluon sector of QCD is the
relatively heavy $\eta'$, which is related to the so-called $U_A(1)$ problem
\cite{weinberg1,weinberg}.
Since the work of t'Hooft \cite{thooft} there is little doubt 
that the large $\eta'$ mass has its origin in the gluon sector of QCD.
The singlet $\eta_0$, which is the main component of the $\eta'$
(apart from admixtures of the octet meson $\eta_8$), is expected to 
couple directly to the gluons via the gluon anomaly and gives an additional
mass term for the $\eta'$.
An important ingredient here is the mixing angle which has been
determined rather accurately by experiments via the analyses of the decays
of the $\eta$ and $\eta'$ \cite{pham,ball} during the last years.

There are different approaches to calculate the mass of the $\eta'$.
In their pioneering works Witten \cite{witten} and
Veneziano and Di Vecchia \cite{veneziano} constructed a meson Lagrangian 
which includes the gluon anomaly.
As a byproduct Witten derived a formula which relates the mass of the
$\eta_0$ to the topological susceptibility. Witten's formula
has been a key tools for
theoretical investigations using rather detailed models of the QCD gluon
vacuum such as the instanton model \cite{novikov,schaefer} or the monopole 
condensate model \cite{ezawa}.  
Recently Hutter \cite{hutter} was able to relate the topological 
susceptibility to the gluon condensate in the simple picture of the gluon 
vacuum as an ensemble of uncorrelated instantons and anti-instantons
and obtained a good estimate of the mass of the $\eta'$. 

Also recently the model of the squeezed gluon vacuum has been considered
as an interesting alternative \cite{celenza}-\cite{blaschke} to the above 
approaches.
In the present paper we discuss the squeezed vacuum in a simple variant
and apply it to the $U_A(1)$ problem. 
We shall directly relate the anomaly term to the gluon condensate.
Numerical results for the gluon condensate are compared with other values 
given by Shifman, Vainsthein and Zakharov \cite{zakharov}
and also recently by Narison \cite{narison}.

The paper is organized as follows:
In Section 2 we briefly recall the $U_A(1)$ problem and discuss its
connection to the gluon condensate.
In Section 3 the effective low energy meson Lagrangian  
which includes an anomalous gluon term is quoted and 
in Section 4 the Witten formula is discussed.
In Section 5 we outline the squeezed vacuum. 
In Section 6 the mass of the $\eta'$ is related to the squeezed
gluon condensate and numerical results are given.
Finally our conclusions are drawn.

\section{$U_A(1)$ problem and the gluon condensate}

The pseudoscalar mesons have been well understood as irreducible
representations of the flavour SU(3) .
The pions and kaons are members of the octet representation, whereas
the $\eta$ and $\eta'$ mesons are related to the octet and singlet
pseudoscalar states
\begin{eqnarray} 
\eta_8  &=& (u\bar{u}+d\bar{d}-2s\bar{s})/\sqrt{6}\nonumber\\
\eta_0  &=& (u\bar{u}+d\bar{d}+s\bar{s})/\sqrt{3}
\end{eqnarray}
via the mixing
\begin{eqnarray}
\eta &=& \eta_8\cos\phi - \eta_0\sin\phi \nonumber\\
\eta'&=& \eta_8\sin\phi + \eta_0\cos\phi
\end{eqnarray}
with a mixing angle $\phi$.

Different experimental data have been used to determine the 
mixing angle $\phi$.
Recent analyses of $\eta $ and $\eta'$ decays \cite{pham} have
obtained a mixing angle
\begin{eqnarray} 
\label{phiexp}
\phi &=& -(18.4\pm 2)^o~.
\end{eqnarray}
The experimental values for the masses of the $\eta$ and the $\eta'$ are
\cite{partdat}
\begin{eqnarray}
\label{massexp}
m_{\eta} &=& 547.45\pm 0.19~{\rm MeV}~, \nonumber\\ 
m_{\eta'} &=& 957.77\pm 0.14~{\rm MeV}~.
\end{eqnarray}
>From these experimental values of $m_{\eta},m_{\eta'}$ and $\phi$ one obtains
the masses $M_{88}$ and $M_{00}$ of the $\eta_8$ and $\eta_0$
\begin{eqnarray}
M_{88} &=& \sqrt{ m^2_{\eta}\cos^2\phi+m_{\eta'}^2\sin^2\phi}
          = 591.49^{+9.31}_{-8.58} ~{\rm MeV}~,\label{M88} \\
M_{00} &=& \sqrt{m^2_{\eta'}\cos^2\phi+m_{\eta}^2\sin^2\phi} 
          = 931.21^{-5.94}_{+5.40} ~{\rm MeV}~.\label{M00}
\end{eqnarray}

On the other hand the masses $M_{88}$ and $M_{00}$ can be obtained from
the following Gell-Mann--Okubo formulae
 \cite{weinberg} 
\begin{eqnarray}
\label{GMO1}
M_{88}^2\big|_{\rm quark} &=&
{1\over 3}\left(2m_{K^+}^2+2m_{K^0}^2-2m^2_{\pi^+}+m^2_{\pi^0}\right)~,\\
\label{GMO2}
M_{00}^2\big|_{\rm quark} &=&
{f_{\pi}^2\over 3f_0^2}
\left(m_{K^+}^2+m_{K^0}^2+m^2_{\pi^+}\right)~,
\end{eqnarray}
which have been derived using the Gell-Mann--Oakes--Renner relations 
for the nonet of pseudoscalar Goldstone bosons under the assumption of only 
small explicit chiral $SU(3)$ symmetry
breaking by the $u$, $d$ and $s$ quark masses.
Here $f_{\pi}=93$ MeV is the pion decay constant and $f_0 \sim f_{\pi}$
\cite{ball} is the singlet decay constant.
Taking the kaon and pion masses from experiment, this relation gives
$M_{88}\big|_{\rm quark}= 566~{\rm MeV}$, 
which is in reasonable agreement with (\ref{M88})\footnote{ 
A value of $\phi=-10.1^o$ would lead to complete agreement
\protect{\cite{partdat}}.
This deviation from the experimental value (\ref{phiexp}) could have its 
origin in a small deviation from Gell-Mann--Okubo formula which can lead to 
a relatively large change in the mixing angle.}.
For the $\eta_0$ mass, however, the Gell-Mann--Okubo formula yields only
a value $M_{00}\big|_{\rm quark}= 413~{\rm MeV}$ for $f_0=f_{\pi}$,
which is much smaller 
than the experimental (\ref{M00}). The large squared difference
\begin{eqnarray}
\label{Dm}
\Delta m_{\eta_0}^2 &=& M_{00}^2-M_{00}^2\big|_{\rm quark}
=0.696 ~{\rm GeV^2}
\end{eqnarray}
shows that the small explicit chiral symmetry breaking by the current
quark masses alone cannot account for the large mass of the $\eta_0$
or the $\eta'$.
This constitutes the well-known $U_A(1)$ problem \cite{weinberg1}.

On the way towards the solution of this puzzle it is important to note
that even in the chiral limit of vanishing quark masses 
the colour singlet axial $U(1)$ quark current 
$j^{\mu}_5\equiv i\bar{q}\gamma^{\mu}\gamma_5 q$ is actually not conserved 
on the quantum level
but afflicted with an anomaly due to the gluon sector of QCD 
\cite{shifman1}
\begin{eqnarray}
\label{anomaly}  
\partial_{\mu}j_5^{\mu} &=& 
2i\bar{q}\gamma_5M_q q+2N_f Q(x)
\end{eqnarray}
with 
\begin{eqnarray}
\label{CPd}
 Q(x) &\equiv & {\alpha_s\over 8\pi} 
G^{\mu\nu a}(x)\tilde{G}_{\mu\nu}^{a}(x)\ ~,\ \ \
\tilde{G}^{\mu\nu a}\equiv 
{1\over 2}\epsilon^{\mu\nu\sigma\rho}G_{\rho\sigma}^{a}~,
\end{eqnarray}
where $G^a_{\mu\nu}(x)$ is the gluon field strength tensor, 
$M_q={\rm diag}(m_u,m_d,m_s)$ is the diagonal matrix of the current quark 
masses, $\alpha_s\equiv g^2/4\pi$ the strong coupling constant and $N_f=3$ 
the number of light quark flavours.
Although derived only in the one-loop approximation in the presence of 
classical background gluon fields it is generally accepted 
that (\ref{anomaly}) is actually an operator identity
(see e.g. the discussion in \cite{shifman1}).
The anomalous term $Q(x)$ is known as Pontryagin density.
An important condition for a large $\eta'$ mass is that there is a
possibility for a nonvanishing $Q(x)$ as will be discussed in the following. 

An important concept in this context is the gluon condensate, defined as
the expectation value of the local gluonic operator 
\begin{eqnarray}
\label{GCop}
N(x) &\equiv & \alpha_s G_{\mu\nu}^{a}(x)G^{\mu\nu a}(x)
\end{eqnarray} 
in the nonperturbative QCD vacuum \cite{shifman1}
\begin{eqnarray}
\label{glc}
\langle{}\alpha_sG^2\rangle ~
&\equiv & \langle{}\alpha_sG_{\mu\nu}^{a}(0)G^{\mu\nu a}(0)\rangle ~
\equiv \langle{}N(0)\rangle ~.
\end{eqnarray}

Approximate empirical values for the physical gluon condensate
are the estimate
$\langle{}\alpha_s G^2\rangle ~\simeq 0.04~ {\rm GeV^4}$ 
by Shifman, Vainshtein and Zakharov \cite{zakharov}
and the update average value 
$\langle{}\alpha_s G^2\rangle ~=(0.071\pm 0.009)~ {\rm GeV^4}$ 
obtained by Narison \cite{narison} in a recent
analysis of heavy quarkonia mass-splittings in QCD.
The value of $\alpha_s$ in the low energy region is not known
very well from experiment.
The value used by Shifman, Vainshtein and 
Zakharov \cite{zakharov} is $\alpha_s\approx 1$ and that used
by Narison \cite{narison} in the low energy region is
$\alpha_s(1.3~ {\rm GeV})\simeq 0.64^{+0.36}_{-0.18}\pm 0.02$. 

A nonvanishing value of the gluon condensate can
lead to a nonvanishing anomalous density $Q(x)$ and therefore
to a large mass of the $\eta'$.
As discussed by Hutter \cite{hutter} for a dilute noninteracting 
gas of statistically independent instantons and anti-instantons ,
which are (anti-)selfdual field configurations in Euclidean space with
$\tilde{G}_{\mu\nu}^a=\pm G_{\mu\nu}^a$, 
the local gluonic operator $N(x)$ defined in
(\ref{GCop}) is proportional 
to the sum of the number densities of instantons and anti-instantons. 
The Pontryagin density $Q(x)$ given in (\ref{CPd}), 
on the other hand, is equal to the
difference of the number densities of instantons and anti-instantons
in the dilute instanton gas. 
In the model of the QCD vacuum as a non-interacting
ensemble of instantons and anti-instantons the existence of a gluon condensate
is therefore a necessary condition for a nonvanishing value of $Q(x)$ and 
hence for a large mass of the $\eta'$.
Furthermore it has been suggested in \cite{blaschke} and will be discussed
in more detail in this paper that also a squeezed condensate in Minkowski
space can lead to a large mass of the $\eta'$ through the $U_A(1)$ anomaly. 

There are several ways to implement the gluon anomaly in calculations
of the meson spectrum in order to obtain
the large value of the $\eta'$ mass.
In Ref. \cite{thooft} t'Hooft introduced
an effective quark interaction in Minkowski space
 simulating the anomalous term  
which breaks $U_A(1)$ but conserves
the chiral $SU(3)_L\otimes SU(3)_R$ symmetry. This determinantal interaction
has been widely used within effective quark models such as the NJL model
\cite{weise,dmitra}.
Dorokhov and Kochelev \cite{dorokhov} model the $\eta'$ as a MIT bag where
the nonperturbative vacuum of QCD is allowed to enter the bag which leads
to instanton induced quark interactions. They obtain the values 
$m_{\eta}= 750 ~{\rm MeV}$ and $m_{\eta'}= 1150 ~{\rm MeV}$. The pions 
obtained in the same scenario, however, turn out to be much too heavy.
For the calculation of the mass of the $\eta'$ mass in microscopic models
of the gluon vacuum, 
such as the magnetic monopole condensate, the instanton gas model or the
squeezed condensate, another
quite general approach is very convenient and will be discussed in the next 
two sections. It is based on Witten's formula
derived from a low energy meson Lagrangian which contains the axial $U_A(1)$
anomaly at tree level.

\section{Effective low energy meson Lagrangian including the chiral anomaly}

In a quite general framework, without using
the concept of instantons, the authors  
\cite{crew}-\cite{volkov}  start from the low energy effective glueball-meson
Lagrangian in a general $\theta$ vacuum
\begin{eqnarray}
\label{U}
{\cal L}^{\rm meson}(U,{\cal Q})= & & 
-\frac{f_{\pi}^2}{4}{\rm Tr} [\partial_{\mu} U\partial^{\mu} U^+]+
\frac{1}{2}v {\rm Tr} [M_q(U^+ +U)] \nonumber\\
&+&\frac{i}{2}{\cal Q}{\rm Tr} [\ln U -\ln U^+]
+\frac{N_f}{af_0^2}{\cal Q}^2 -\theta {\cal Q}~,
\end{eqnarray}
where
\begin{eqnarray}
U(x) &\equiv & \exp \left[i\frac{\sqrt{2}}{f_{\pi}}
\left(\sum_{a=1}^8 B_a(x)\lambda_a+{f_{\pi}\over \sqrt{3}f_0}\eta_0(x)
1_3\right)
\right]~,
\end{eqnarray}
with the octet meson fields $B_a$ 
\begin{eqnarray}
\sum_{a=1}^8 B_a\lambda_a &=& 
\left[\begin{array}{ccc}
{1\over \sqrt{2}}\pi^0+{1\over \sqrt{6}}\eta_8 & \pi^+ & K^{+} \\
\pi^- &  -{1\over \sqrt{2}}\pi^0+{1\over \sqrt{6}}\eta_8 & K^0 \\
 \bar{K}^- &\bar{K}^0 & -\sqrt{2\over 3}\eta_8 
\end{array}\right]
\end{eqnarray}
and the singlet $\eta_0$. The pseudoscalar glueball field ${\cal Q}(x)$
plays the role of an auxilary field included
into the meson Lagrangian in order to implement the axial $U_A(1)$ anomaly
on the hadronic level. It can be identified with the vacuum expectation value
of the Pontryagin density $Q(x)$ defined in (\ref{CPd}) in terms of gluon
degrees of freedom as will be  discussed further below.
The vacuum angle $\theta$, the parameters $a$ and $v$, as well as
the diagonal mass matrix of the current quarks $M_q= {\rm diag}(m_u,m_d,m_s)$ 
are to be fixed by comparison with experiment.
Measurements of the dipole moment of the neutron \cite{cheng} limit the vacuum 
angle to $\theta < 10^{-9}$, so that in practice it can be taken equal 
to zero.

The kinetic term reads explicitly
\begin{eqnarray}
-\frac{f_{\pi}^2}{4}{\rm Tr} [\partial_{\mu} U\partial^{\mu} U^+] &=&
-{1\over 2}\partial_{\mu}\pi^0\partial^{\mu}\pi^0
-\partial_{\mu}\pi^+\partial^{\mu}\pi^-
-\partial_{\mu}K^+\partial^{\mu}K^- \nonumber\\
& &
-\partial_{\mu}K^0\partial^{\mu}\bar{K}^0
-{1\over 2}\partial_{\mu}\eta_8\partial^{\mu}\eta_8
-{1\over 2}\partial_{\mu}\eta_0\partial^{\mu}\eta_0
\end{eqnarray}
in terms of the pion, kaon and eta fields.
The explicitly chiral symmetry breaking mass term is
\begin{eqnarray}
\frac{1}{2}v {\rm Tr} [M_q(U^+ +U)] &=& {1\over 2} v\Bigg[
{1\over 2}(m_u+m_d)(\pi^0\pi^0 +2\pi^+\pi^-) \nonumber\\
& &+(m_u+m_s)K^+\bar{K}^- +(m_d+m_s)K^0\bar{K}^0 \nonumber\\
& &+{1\over 6}(m_u+m_d+4m_s)\eta_8^2
   +{f_{\pi}^2\over 3 f_0^2}(m_u+m_d+m_s)\eta_0^2\nonumber\\
& &+(m_u-m_d){1\over \sqrt{3}}\pi^0 
    (\eta_8 + {\sqrt{2}f_{\pi}\over f_0}\eta_0)\nonumber\\
& &+{\sqrt{2}f_{\pi}\over 3f_0}(m_u+m_d-2m_s)\eta_8\eta_0
\Bigg]~.
\label{mesonmasses}
\end{eqnarray}
The corresponding meson mass formulae implicit in (\ref{mesonmasses})
are the Gell-Mann--Oakes--Renner relations. They can be combined in order to 
yield the Gell-Mann--Okubo relations (\ref{GMO1}) and (\ref{GMO2}).

Since
\begin{eqnarray}
\label{eta_0U} 
-{i\over 2}{\rm Tr}[\ln U -\ln U^+] &=& {\sqrt{2N_f}\over f_0}
\eta_0~,
\end{eqnarray}
only the singlet field $\eta_0$ is coupled to ${\cal Q}(x)$.
The Euler-Lagrange equations for $U$ and $U^+$ of the meson Lagrangian
(\ref{U}) include the chiral anomaly
\begin{eqnarray}
\label{anomalyb} 
\partial_{\mu}A_0^{\mu} &=& -iv{\rm Tr}[M_q(U^+-U)]
+2N_f {\cal Q}(x)
\end{eqnarray}
with the axial U(1) current 
$A_0^{\mu}=-i{f_{\pi}^3\over 2 f_0}
{\rm Tr}[U\partial_{\mu}U^+ -U^+\partial_{\mu}U]$ from
Noethers theorem.
Eq. (\ref{anomalyb}) is the hadron analogue to (\ref{anomaly})
with the pseudoscalar glueball field
identified as the vacuum expectation value
of the Pontryagin density $Q(x)$ defined in (\ref{CPd}) in terms of gluon
degrees of freedom. 
The Lagrangian (\ref{U}) therefore includes the chiral anomaly.

Using $\delta {\cal L}^{\rm meson}/\delta {\cal Q}=0$ to eliminate the 
auxilary pseudoscalar glueball field ${\cal Q}(x)$
leads to the reduced Lagrangian
\begin{eqnarray}
\label{Ured}
{\cal L}^{\rm meson}_{\rm red}(U)&=& 
-\frac{f_{\pi}^2}{4}{\rm Tr}[\partial_\mu U\partial^\mu U^+]+
\frac{1}{2}v{\rm Tr}[M_q(U^+ +U)] \nonumber\\
& &-\frac{af_0^2}{4 N_f}\left(\theta-\frac{i}{2}{\rm Tr}
[\ln U -\ln U^+]\right)^2~.
\end{eqnarray}
Noting (\ref{eta_0U}) we see that the $\eta_0$ field  attains an additional 
contribution 
\begin{eqnarray}
\label{Dma}
\Delta m_{\eta_0}^2 &=& a
\end{eqnarray}
to its mass from the gluon anomaly, which can be
chosen in accordance with (\ref{Dm}).
A different derivation of the reduced Lagrangian 
(\ref{Ured}), which contains the $U_A(1)$ anomaly at tree level,
 has been given by Witten \cite{witten} using large $N_c$ arguments. 
Note that the kinetic term and the mass term in the Lagrangian
(\ref{Ured}) are of order $O(1)$ in the number $N_c$ of colours,
since $f_0\sim f_{\pi}\sim O(N_c^{1/2})$, whereas the 
third anomalous term like
the mass $a$ due to the gluon anomaly itself are of order $O(N_c^{-1})$
and therefore vanish in the large $N_c$ limit.

\section{Witten's formula for the $\eta_0$ mass}

>From the anomalous low energy meson Lagrangian (\ref{Ured}) one can 
derive the quite general formula by Witten \cite{witten}, which allows one
to calulate the mass shift of the $\eta_0$ due to the gluon anomaly
in leading order in $1/N_c$ for microscopic models of the gluon vacuum.
Although $\theta$ is practically zero, Witten proposed to use its fluctuations
to calculate the quadratic mass shift $\Delta m_{\eta_0}^2$ in the following
way. From (\ref{Ured}) and (\ref{Dma}) one obtains
\begin{eqnarray}
\label{etam}
\Delta m^2_{\eta_0} &=& {2N_f\over f_{\pi}^2}
\left({d^2\varepsilon_0\over d\theta^2}
\right)_{\theta=0}^{\rm no\ quarks}~.
\end{eqnarray}
Here $\varepsilon_0(\theta)=E_0(\theta)/V$ is the ground state 
energy density of the Hamiltonian corresponding
to (\ref{Ured}) with no quarks, i.e. all meson fields set equal to zero.
The singlet decay constant $f_0$ has been replaced by $f_{\pi}$ which
is in agreement with experiment \cite{ball} and in accordance with large $N_c$
arguments. 
As noted by Witten, the ground state energy $E_0(\theta)$ 
can be alternatively considered
in pure gluon QCD with the Lagrangian including an anomalous term as
\begin{eqnarray}
{\cal L}^{\rm gluon} &=& -{1\over 4}G^{\mu\nu}_a G_{\mu\nu}^a + 
\frac{\alpha_s\theta}{8\pi}G^{\mu\nu}_a \tilde{G}_{\mu\nu}^a
\end{eqnarray}
with
\begin{eqnarray} 
G^{\mu\nu a} G_{\mu\nu}^a &=& -2 (E_{i}^a)^2 + 2 (B_i^a)^2 ~~,~~
G^{\mu\nu a} \tilde{G}_{\mu\nu}^a=- 4 E_{i}^a B_i^a~~.
\end{eqnarray}
$E_i^a$ and $B_i^a$ are the components of the chromoelectric and
chromomagnetic field strength.
For the quantization in the Hamilton formalism
we use the Weyl gauge $A_0=0$, such that $E_{i}^a=\dot{A}_i^a$.
Introducing the canonical momenta 
\begin{eqnarray}
\Pi_i^a &\equiv & 
\frac{\partial{\cal L}_{\rm singlet}}
{\partial\dot {A}_i^a}=E_{i}^a+{\alpha_s \theta\over 2\pi}B_i^a~,  
\end{eqnarray} 
the Hamiltonian reads
\begin{eqnarray}
\label{htheta}
H_{\rm gluon} &=&
\frac{1}{2}\int d^3\vec{x}\left[
\left(\Pi_i^a-\frac{\alpha_s\theta}{2\pi}B_i^a\right)^2
+(B_i^a)^2\right]~.
\end{eqnarray} 
The Hamiltonian contains a term linear in $\theta$
\begin{eqnarray}
I_1 &=& -2\int d^3\vec{x}~\frac{\alpha_s\theta}{4\pi}\Pi_i^a B_i^a
\end{eqnarray} 
and a term quadratic in $\theta$
\begin{eqnarray}
I_2 &=& 2\int d^3\vec{x}\left(\frac{\alpha_s\theta}{4\pi}\right)^2 (B_i^a)^2~.
\end{eqnarray} 
In order to calculate the topological susceptibility 
$\left(d^2\varepsilon_0/d\theta^2\right)_{\theta = 0}$ 
it is only necessary to calculate 
the vacuum energy $E_0(\theta)$ up to second order in $\theta$. 
For this purpose one has to do second order perturbation theory 
in the operator $I_1$ and only first order in $I_2$.
This way Witten \cite{witten}  derived an expression for
the topological susceptibility of QCD without quarks:
\begin{eqnarray}
\label{witfo}
\left({d^2\varepsilon_0\over d\theta^2}\right)^{\rm no ~quarks}_{\theta=0} &=& 
-i\int dtd^3\vec{x} 
~\langle{}0_I|{\cal T}Q(\vec{x},t)Q(\vec{0},0)
|0_I\rangle _{\rm conn}
\nonumber\\
&+& 4\left(\frac{\alpha_s}{4\pi}\right)^2 
~\langle{}0_I|B_i^a(\vec{0},0)^2|0_I\rangle ~.
\end{eqnarray}
where $|0_I\rangle $ is the interaction picture
gluon vacuum corresponding to $\theta = 0$.   
The subscript ''conn'' denotes the connected part of the Green function
and ${\cal T}$ is the Dyson time ordering operator.
Eqs. (\ref{etam}) and (\ref{witfo}) together constitute the 
Witten formula. 
The first term in (\ref{witfo}) is a propagator term whereas the second one
is a contact term. The contact term can be incorporated into the propagator 
term, if instead of the Dyson ${\cal T}$ ordering the Wick 
${\cal T}^*$
ordering is used \cite{diakon}. In the following we shall use (\ref{witfo})
with the Dyson ${\cal T}$ ordering. Witten kept only the propagator term
in (\ref{witfo}) and dropped the contact term. 
He argued that although the contact term is
necessary for the positivity of the result, it should not contribute
to the mass of the $\eta'$ since no Goldstone pole
can appear in a one point function.
Based on Witten's formulae (\ref{etam}) and (\ref{witfo}), 
dropping the contact term, there were several approaches to calculate 
the large mass of the $\eta'$.

As one of the earliest approaches Novikov et al. \cite{novikov}
used the Euclidean model of the gluon vacuum as an ensemble of 
noninteracting (anti-)instantons. In this context is valid 
\begin{eqnarray}
\label{QQNN}
\langle{}0_{\rm ig}|{\cal T}Q(x)Q(0)|0_{\rm ig}\rangle _{\rm conn} &=&
{1\over 64\pi^2}\langle{}0_{\rm ig}|{\cal T}N(x)N(0)|0_{\rm ig}\rangle 
_{\rm conn}~ 
\end{eqnarray}
with $Q(x)$ and $N(x)$ given in (\ref{CPd}) and
(\ref{GCop}) and  $|0_{\rm ig}\rangle $ denoting the dilute 
instanton gas vacuum.
He then related the right hand side of (\ref{QQNN}) to the gluon condensate
using the Ward identity \cite{shifman1}
\begin{eqnarray}
\label{NNN}
\int d^4x{1\over 8\pi}\langle{}0|{\cal T}N(x)N(0)
|0\rangle _{\rm conn}
&=& {12\over 11 N_c}\langle{}0|N(0)|0\rangle
\end{eqnarray}
corresponding to the anomalous breaking of scale invariance. 
Inserting (\ref{QQNN}) and (\ref{NNN}) into  Witten's 
formulae (\ref{etam})  and (\ref{witfo}) in Euclidean space, he obtained
the following relation between the mass of the $\eta'$ and the physical value 
of the gluon condensate $\langle{}\alpha_s G^2\rangle$
\begin{eqnarray}
\label{metanov}
\Delta m_{\eta_0}^2 &=& {2N_f\over 8\pi f_{\pi}^2}{12\over 11 N_c}
{1\over \gamma}\langle{}\alpha_s G^2\rangle~.
\end{eqnarray}
The coefficient $\gamma < ~ 1$, originating from the fermion determinant 
in the QCD path integral, accounts for the suppression of
the gluon condensate value due to the presence of light quarks  
\begin{eqnarray}
\label{lq}
\langle{}\alpha_s G^2\rangle &=&
\gamma \langle{}\alpha_s G^2\rangle_{\rm no ~quarks}~.
\end{eqnarray}
A discussion of the value for $\gamma$ has been given by Novikov et al.
\cite{novikov} in the instanton gas scenario. They find values in the range
$\gamma\simeq 1/3 - 1/2$.  
Recently Hutter \cite{hutter} showed that formula (\ref{metanov}) for the mass
of the $\eta'$ remains valid even for the more general case of the gluon 
vacuum as an ensemble of noninteracting instantons and anti-instantons.
Using the somewhat wider range $\gamma\simeq 0.4 -0.7$, 
Hutter obtains a value $m_{\eta'} = 884 \pm 116 ~{\rm MeV}$ 
neglecting the quark masses.

Another application of the above Witten formula 
has been the calculation of the mass 
of the $\eta'$ in the magnetic monopole condensate scenario by
Ezawa and Iwazaki \cite{ezawa}. Using the hypothesis of dominance of
the Abelian gauge field components at large distances they find a value
$m_{\eta'}= 550 ~{\rm MeV}$ neglecting
the influence of quarks on the gluon condensate.

A further interesting possibility to explain the large
mass of the $\eta'$, which has been suggested recently \cite{blaschke}
is the model of a squeezed gluon vacuum
to be discussed in the following two sections.

\section{The model of the squeezed gluon vacuum}

The squeezed condensate of gluons has been investigated recently
\cite{celenza}-\cite{blaschke} in order to construct a Lorentz and gauge
invariant stable QCD vacuum in Minkowski space. 
Different alternative approaches
have not solved this problem. For instance the simple perturbative vacuum 
is unstable \cite{sav}, and there is no stable 
(gauge invariant) coherent vacuum in Minkowski space \cite{leut}.
>From the physical point of view, the squeezed state differs from the coherent 
one by the condensation of colour singlet gluon pairs rather than of single 
gluons. In analogy to the Bogoliubov model \cite{nn}
we consider the case of a homogeneous condensate, but
in a squeezed instead of a coherent state.

Let $|n\rangle $ denote the eigenstates of the pure gluon 
Hamiltonian (\ref{htheta}) for vanishing vacuum angle $\theta=0$.
In order to introduce the squeezed states let us consider the gluon system
to be enclosed in a large finite volume $V$.
The squeezed states $|n_{\rm sq}[\xi]\rangle ~$ as candidates for
the gluon eigenstates $|n\rangle $, in particular 
the squeezed vacuum $|0_{\rm sq}[\xi]\rangle ~$ 
as a candidate for a homogeneous colourless gluon vacuum 
$|0\rangle $, 
are constructed from the nonperturbative states 
$|n^{(0)}\rangle ~~\equiv |n_{\rm sq}[\xi]\rangle\big|_{\xi=0}$, 
further specified below, according to
\begin{eqnarray}
\label{sqt1}
|n_{\rm sq}[\xi]\rangle ~\ &=&\ U_{\rm sq}^{-1}[\xi]|n^{(0)}\rangle ~~.
\end{eqnarray}
The squeezing operator
\begin{eqnarray}
\label{sqt2}
U_{\rm sq}[\xi] &=& \exp\left[i{\xi\over 2}V({\cal A}^a_{i}
{\cal E}^a_{i}+
{\cal E}^a_{i}{\cal A}^a_{i})\right]
\end{eqnarray}
with the zero momentum components ${\cal A}^a_{i}$ and ${\cal E}^a_{i}$ 
of the fields and their canonical momenta contains the  parameter $\xi$ 
given below.
This special transformation for the homogeneous condensate
does not violate Lorentz invariance,
since the gauge fields are massless \cite{Linde}.
The question of gauge invariance of such a procedure is a difficult
open problem and first steps towards a clarification are 
under current investigation \cite{khvede}. 
As in Ref. \cite{schuette}
we suppose here the gauge invariance of ${\cal A}_i^a{\cal E}_i^a$ and 
hence of the squeezing operator as a colourless functional of the spatial 
zero momentum components of the gauge fields.
The multiplicative transformations of fields corresponding to
(\ref{sqt1}) and (\ref{sqt2}) are
\begin{eqnarray}
\label{usq}
U_{\rm sq}[\xi]~{\cal A}_i^a~U_{\rm sq}^{-1}[\xi] &=&
 {\rm e}^{\xi} {\cal A}_i^a~,
\nonumber\\ \vspace{0.2cm}
U_{\rm sq}[\xi]~{\cal E}_i^a~U_{\rm sq}^{-1}[\xi] &=&
 {\rm e}^{ - \xi} {\cal E}_i^a~.
\end{eqnarray}
>From this canonical transformation it follows that
the expectation values in the squeezed state basis as
functions of the squeezing parameter $\xi$ behave like
\begin{eqnarray} 
\langle{}n_{\rm sq}[\xi]|\left({\cal B}_i^a\right)^2
|n'_{\rm sq}[\xi]\rangle ~ 
&=& {\rm e}^{4\xi}\langle{}n^{(0)}|\left({\cal B}_i^a\right)^2|n'^{(0)}
\rangle ~~,
\label{b2sq}
\end{eqnarray}
\begin{eqnarray}  
\langle{}n_{\rm sq}[\xi]|\left({\cal E}_i^a\right)^2
|n'_{\rm sq}[\xi]\rangle ~ &=& 
 {\rm e}^{-2\xi}\langle{}n^{(0)}|\left({\cal E}_i^a\right)^2
|n'^{(0)}\rangle ~~,\label{e2sq}
\end{eqnarray}
\begin{eqnarray} 
\langle{}n_{\rm sq}[\xi]|{\cal E}_i^a {\cal B}_i^a
|n'_{\rm sq}[\xi]\rangle ~ &=& 
 {\rm e}^{\xi}\langle{}n^{(0)}|{\cal E}_i^a {\cal B}_i^a
|n'^{(0)}\rangle ~~,\label{ebsq}
\end{eqnarray}
with ${\cal B}^a_i\equiv f^{abc}\epsilon_{ijk}
{\cal A}^b_j {\cal A}_k^c$. 
Let the reference states $|n^{(0)}\rangle ~$ be such that the expectation 
values
$\langle{}n^{(0)}|\left({\cal B}_i^a\right)^2|n'^{(0)}\rangle$, 
$\langle{}n^{(0)}|\left({\cal E}_i^a\right)^2|n'^{(0)}\rangle$
and $\langle{}n^{(0)}|{\cal E}_i^a {\cal B}_i^a |n'^{(0)}\rangle$ 
behave in the large volume limit ($V \to \infty$)  like $V^{-4/3}$ in 
accordance with dimensional analysis.
The parameter of the squeezing transformation $\xi$ can be chosen so that
the magnetic condensate density (\ref{b2sq}) remains finite in the large 
volume limit (${\rm e}^{4\xi}\sim V^{4/3}$)
\begin{eqnarray}
\label{B}
\lim_{V \rightarrow \infty} 
\langle{}n_{\rm sq}|\left({\cal B}_i^a\right)^2
|n'_{\rm sq}\rangle ~~ &=& {\cal O} [1]~.
\end{eqnarray}
We shall denote the corresponding squeezed states simply by 
$|n_{\rm sq}\rangle$. 
This entails that the electric component (\ref{e2sq}) and the mixed
component (\ref{ebsq}) of the gluon condensate vanish in the large volume 
limit 
\begin{eqnarray}
\label{E}
\lim_{V \rightarrow \infty} 
\langle{}n_{\rm sq}|\left({\cal E}_i^a\right)^2
|n'_{\rm sq}\rangle ~~ &=& {\cal O} [1/V^2]~,
\end{eqnarray}
\begin{eqnarray}
\label{EB}
\lim_{V \rightarrow \infty} 
\langle{}n_{\rm sq}|{\cal E}_i^a{\cal B}_i^a
|n'_{\rm sq}\rangle ~~ &=&{\cal O} [1/V]~.
\end{eqnarray}
Hence we conclude that in the squeezed vacuum (\ref{sqt1})
the gluon condensate
is equal to its magnetic part,
\begin{eqnarray}
\label{asg2}
\langle{}\alpha_s G^2\rangle ~_{\rm no ~quarks}&=& 
\langle{}0_{\rm sq}|
\alpha_s G^{\mu\nu a}(0)G_{\mu\nu}^a(0)|0_{\rm sq}
\rangle ~ \nonumber\\
&=& 2~ \langle{}0_{\rm sq}|\alpha_s({\cal B}_i^a)^2
|0_{\rm sq}\rangle ~ ~.
\end{eqnarray}

Note that this model is consistent with the picture of the QCD vacuum as a 
homogenous magnetic medium.
As discussed e.g. in \cite{shuryak}, the vacuum energy 
$\varepsilon_{\rm vac}$ of full QCD can be related to the gluon condensate 
$\langle{}\alpha_s G^2\rangle$ via
the one-loop result for the scale anomaly in the dilatation 
current
\begin{eqnarray}
\varepsilon_{\rm vac} &=& {1\over 4}\langle\theta_{\mu}^{\mu}\rangle 
\cong -{b\over 32\pi}\langle{}\alpha_s G^2\rangle~.
\end{eqnarray}
Here $\langle\theta_{\mu}^{\mu}\rangle$ is the
vacuum expectation value of the trace of the energy momentum tensor and 
$b={11\over 3}N_c-{2\over 3}N_f$ the QCD $\beta$-function coefficient.
Since the vacuum energy $\varepsilon_{\rm vac}$ is negative and $b$ positive, 
the gluon condensate $\langle{}\alpha_s G^2\rangle$
is expected to be positive and hence 
\begin{eqnarray}
\label{magnele}
\langle{}\alpha_s B^2\rangle &>& \langle{}\alpha_s E^2\rangle ~,
\end{eqnarray}
which is a Lorentz invariant statement and corresponds to a magnetic vacuum.
Quark condensate terms have been neglected in this estimation and
are expected to soften the inequality (\ref{magnele}) but not to turn it
to the reverse.
The squeezed vacuum has vanishing electric field $E$ and thus is in 
accordance with (\ref{magnele}).

\section{The $\eta'$ mass in the model of the squeezed gluon vacuum}

For the calculation of the $\eta'$ mass in the squeezed vacuum it is useful
to rewrite the Witten formula (\ref{witfo}) in the Schr\"odinger picture as
\cite{diakon}
\begin{eqnarray}
\label{witfoS}
\left({d^2\varepsilon_0\over d\theta^2}
\right)^{\rm no ~quarks}_{\theta=0} &=& 
 -2\sum_{n \neq 0}{|\langle{}n|Q(\vec{0})|0\rangle|^2 \over 
\varepsilon_n-\varepsilon_0}
+ 4\left(\frac{\alpha_s}{4\pi}\right)^2 
\langle{}0|(B_i^a(\vec{0}))^2|0\rangle ~.
\end{eqnarray}
It contains the exact eigenstates $|n\rangle $ and eigenvalues $\epsilon_n$
of the pure gluon QCD Hamiltonian. We approximate (\ref{witfoS}) by
replacing the exact eigenstates and eigenvalues by the nonperturbative
squeezed states and the corresponding energy expectation values.
In order to see whether the propagator and the contact term give 
finite contributions we inspect their volume dependence.
Since in the squeezed vacuum
\begin{eqnarray}
\langle{}0_{\rm sq}|(B_i^a(\vec{0}))^2|0_{\rm sq}\rangle &=&
\langle{}0_{\rm sq}|({\cal B}_i^a)^2|0_{\rm sq}\rangle~, 
\end{eqnarray}
the contact term gives a finite contribution to the topological 
susceptibility according to (\ref{B}).

In addition a further, negativ contribution might arise from the propagator 
term. 
Despite the fact that the matrix elements are suppressed in the large volume 
limit the denominator can simultaneously 
become very small due to states arbitrarily close to the vacuum.
Whereas in the instanton model of the gluon vacuum only the propagator term
is considered, as discussed in Section 4, we shall here not further
investigate the propagator term but consider only the finite contribution
from the contact term.

The contact term by itself gives the following contribution to the 
$\eta_0$ mass via (\ref{etam}) 
\begin{eqnarray}
\label{meta0'}
\Delta m_{\eta_0}^2 \Big|_{\rm contact}&=& \frac{3 \alpha_s}{2\pi^2 f_{\pi}^2}
\langle{}0_{\rm sq}|\alpha_s ({\cal B}_i^a)^2
|0_{\rm sq}\rangle ~.
\end{eqnarray}

Using the expression (\ref{asg2}) for the squeezed gluon condensate and 
relation (\ref{lq}) to account for the suppresion of the physical 
gluon condensate due to the presence of light quarks by a factor 
$\gamma < ~ 1$ we obtain 
\begin{eqnarray}
\label{meta0G}
\langle{}\alpha_s G^2\rangle ~
 &=& \frac{4\gamma \pi^2 f_\pi ^2}{3 \alpha_s}
\Delta m_{\eta_0}^2\Big|_{\rm contact} ~.
\end{eqnarray}
This formula is the main result of our investigation.
It relates the gluon condensate to the $U_A(1)$ breaking contact
contribution to the mass shift of the $\eta_0$.

We point out that Nielsen et al. \cite{nielsen} have derived the same
formula for the shift of the $\eta'$ mass as (\ref{meta0'}) (except for the
factor $\gamma$) using the Cheshire cat principle instead of the Witten
formula.

A main source of uncertainty is the reduction factor of the gluon condensate
due to the presence of light quarks.
We shall use here the estimate $\gamma\simeq 1/3-1/2$
obtained by Novikov et al. \cite{novikov} in the instanton gas scenario.

The value of $\alpha_s$ in the low energy region is not known
very well from experiment.
The value used by Shifman, Vainshtein and 
Zakharov  \cite{zakharov} is $\alpha_s\approx 1$ and that used
by Narison \cite{narison} in the low energy region is
$\alpha_s(1.3~ {\rm GeV})\simeq 0.64^{+0.36}_{-0.18}\pm 0.02$. 
In order to check whether relation (\ref{meta0G}) is in agreement with
empirical data we have plotted in Fig. 1 the gluon condensate 
$\langle{}\alpha_s G^2\rangle ~$ 
against $\alpha_s$ for two limiting values $1/3$ and $1/2$ of $\gamma$.
We estimate the contact contribution by the full empirical mass shift 
$\Delta m_{\eta_0}^2= 0.696 \pm 0.02 ~GeV^2$ obtained in (\ref{Dm}) for
$\phi=-18.4^o$ . 
\begin{figure}
\centerline{\psfig{figure=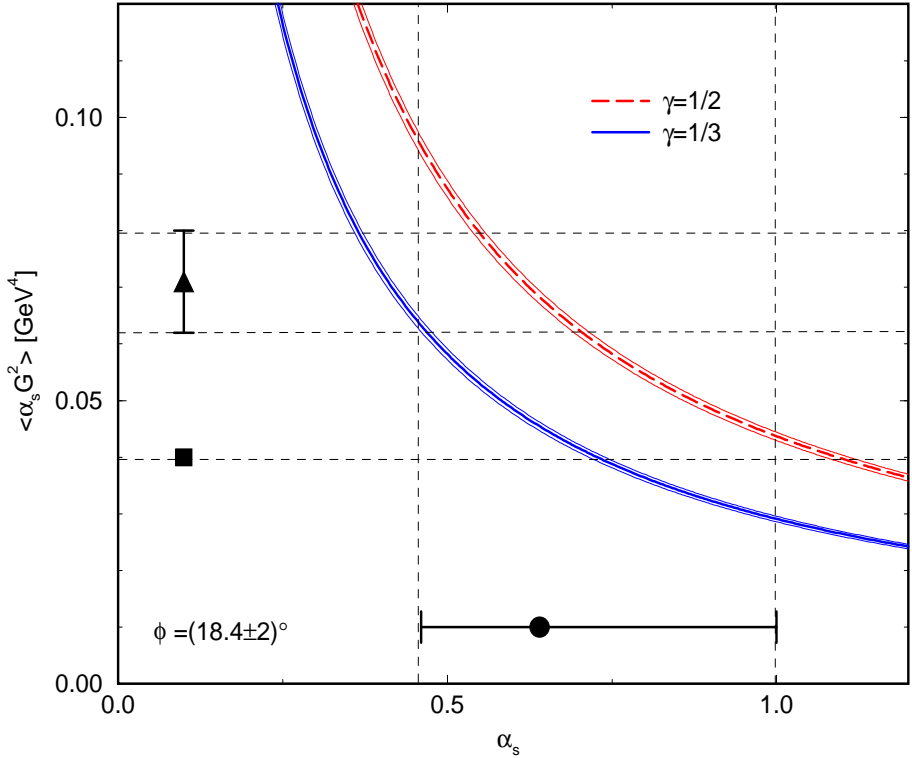,width=14cm,height=12cm,angle=-90}}
\caption{The gluon condensate $\langle{}\alpha_s G^2\rangle ~$ 
vs. QCD coupling $\alpha_s$  according to (\protect\ref{meta0G}) 
and (\protect\ref{Dm}) for the two limiting values 1/3 (solid line) and 
1/2 (dashed line) of the suppression factor $\gamma$
for the mixing angle $\phi=-18.4^o$. 
The thin lines indicate the error $\pm 2^o$ in the value of the mixing angle.
Also shown are the gluon condensate values obtained by Narison 
\protect\cite{narison} (filled triangle) and by Shifman, Vainshtein and 
Zakharov \protect\cite{zakharov} (filled square) which both are compatible 
with the $\alpha_s(1.3~ {\rm GeV})$ value (filled circle) of Ref.
\protect\cite{narison} according to relation (\protect\ref{meta0G}).} 
\end{figure}
The curves in Fig. 1 give our result for these two limiting 
values $\gamma=1/3$ (solid line) and $\gamma=1/2$ (dashed line) 
and the mixing angle
$\phi=-18.4^o$. The thin lines indicate the error $\pm 2^o$ in the value of 
the mixing angle which is negligibly small.
Also shown are the gluon condensate value
$\langle{}\alpha_s G^2\rangle ~\simeq 0.04~ {\rm GeV^4}$ 
by Shifman, Vainshtein and 
Zakharov \cite{zakharov} (filled square)
and the update average value 
$\langle{}\alpha_s G^2\rangle ~=(0.071\pm 0.009)~ {\rm GeV^4}$ 
for the gluon condensate 
obtained by Narison \cite{narison} (filled triangle) in a recent
analysis of heavy quarkonia mass-splittings in QCD.
The gluon condensate values described by our result (\ref{meta0G})  
are in good agreement with both the Shifman, Vainshtein and 
Zakharov  and the Narison value
for the respective values of $\alpha_s$ in the range of Narison's update
average $\alpha_s$ (filled circle).
Compared with our previous result \cite{blaschke} we 
have included in this work a rather detailed discussion of the error 
due to the experimental uncertainty in the mixing angle and 
due to the influence of the quarks on the gluon condensate.

\section{Conclusions}

In the present work we have pointed out the
possibility to resolve the $U_A(1)$ problem
via the model of a homogeneous squeezed gluon condensate
as an interesting alternative to existing models of the QCD gluon vacuum
such as the instanton gas model.
In particular we have discussed that in the squeezed vacuum 
the contact term in Witten's formula can give a sizable contribution
to the gluonic part of the $\eta_0$ mass.  
In the framework of a homogeneous squeezed vacuum we have obtained a relation 
between the value of the gluon condensate and the mass shift of the $\eta_0$
as a function of the strong coupling constant. 
An interesting aspect to consider the contact term is the fact that
H.B. Nielsen et al. \cite{nielsen} have derived exactly the same relation
using the Cheshire cat principle istead of Witten's formula.
The gluon condensate values found in our estimate 
are in quite good agreement with both the 
``standard'' value $0.04~ {\rm GeV}^4$ by Shifman, Vainsthein and Zakharov 
and the update average value $0.071~ {\rm GeV}^4$ by Narison for reasonable 
values of the strong coupling in the low energy region.

In our simple model of the squeezed vacuum
only the zero momentum mode of the gluon field operators has been squeezed.
This is a Lorentz invariant operation since the gauge field is massless.
The question of the gauge invariance of the procedure is still an open problem
and under current investigation \cite{khvede} as well as 
the possibility of a Lorentz
and gauge invariant extension to the squeezing of nonzero momentum modes.

\section*{Acknowledgement}
We thank D. Ebert, I.I. Kogan, E.A. Kuraev  and C.D. Roberts 
for fruitful discussions on the subject.
H.-P.P. is grateful to the Deutsche Forschungsgemeinschaft for support under 
contract No. RO 905/11-1.
M.K.V. acknowledges financial support provided by INTAS under Grant No.
W 94-2915 and by the Max-Planck Gesellschaft as well as the hospitality of the
Fachbereich Physik at the University of Rostock.

\end{document}